\documentclass[number,preprint,12pt]{elsarticle}
\PassOptionsToPackage{hyphens}{url}

\usepackage{amssymb}
\usepackage{amsmath}
\usepackage{tabularx}
\usepackage{longtable}

\usepackage{tikz}
\usepackage{hyperref}
\usetikzlibrary{calc}
\usepackage[font=footnotesize]{caption}
\begin{document}

\hypersetup{pageanchor=false}
\begin{titlepage}
    \begin{center}
        \vspace*{0.1cm}
        {\LARGE AI-Simulated Expert Panels for Socio-Technical Scenarios and Decision Guidance\par}
        \vspace{1.0cm}

                Andrew G. Ross $^{1,*}$  and Alan M. Ross $^{2}$ \\
        \vspace*{0.70\baselineskip}
         $^1$ Forschungszentrum J\"ulich GmbH, Institute of Climate and Energy Systems, J\"ulich Systems Analysis, Wilhelm-Johnen-Strasse, 52425 J\"ulich, Germany. \url{a.ross@fz-juelich.de}\\
        \vspace*{0.5\baselineskip}
         $^2$ Google Core, California, US. \url{alanmross@google.com} \\
        \vspace*{1\baselineskip}
\vspace*{4\baselineskip}
$^*$ Corresponding author: \url{a.ross@fz-juelich.de} \\
\vspace*{4\baselineskip}
\today \\
\vspace*{4\baselineskip}

\textbf{\textcolor{red}{\MakeUppercase{Not peer-reviewed version}}}

\vspace*{2.25\baselineskip}

    \end{center}
\end{titlepage}

\begin{titlepage}
    \begin{center}
        \vspace*{0.1cm}
        {\LARGE AI-Simulated Expert Panels for Socio-Technical Scenarios and Decision Guidance\par}
        \vspace{0.5cm}
    \end{center}

    \vspace{0.5cm}

    {\noindent Socio-technical scenarios for net-zero and other transformation pathways combine qualitative storylines with quantitative models, embedding them in plausible societal contexts for model assessment. Conventional scenario generation is resource-intensive, can be limited in internal consistency and diversity of expert and stakeholder perspectives, and is rarely stress-tested. This paper introduces a synthetic, AI-based expert panel to address these bottlenecks. An AI model first simulates domain experts who agree on descriptors, states, and their interactions. A probabilistic Cross-Impact Balance analysis then generates internally consistent pathways, using stochastic shocks to assess robustness and pathway diversity. An AI stakeholder panel uses multi-criteria decision analysis to select a preferred pathway; an AI expert panel translates it into model-ready quantitative inputs. Although scalable and applicable to any other country or region, the framework is applied to Germany's energy transition as a proof of concept, and offers an alternative and/or supplement to scenario generation. Furthermore, it enables Virtual AI-Led Decision Laboratories for exploratory policy stress-testing and provides an approach for rapid, structured expert elicitation and decision support in other domains.} \\

    \vspace{0.5em}


    \vspace{0.5em}


    \vfill

\end{titlepage}
\hypersetup{pageanchor=true}

\newpage 

\section{Introduction}
\label{sec1}

\noindent Socio-technical scenarios for energy, climate, and related transformation analysis combine qualitative storylines about possible societal and policy futures with quantitative modelling \cite{WEIMERJEHLE2006334,Pregger2020}. This combination embeds future pathways within plausible societal contexts \cite{Weimer-Jehle2020}, giving quantitative analysis its contextual meaning and policy relevance and making the assumptions behind model results explicit for decision-makers \cite{Schweizer2020}. The Cross-Impact Balance (CIB) method offers a formal framework for building these foresight storylines \cite{WEIMERJEHLE2006334}. Experts define descriptors and their alternative future states, elicit cross-impacts in a cross-impact matrix (CIM), run the CIB consistency analysis, select or prioritise scenarios (e.g.\ via multi-criteria decision analysis (MCDA)), and convert the chosen storylines into quantitative assumptions for sectoral models \cite{WEIMERJEHLE2006334,Schweizer2020,KURNIAWAN2022102815,Pregger2020,Prehofer2021,Weimer-Jehle2020}. The process is time- and resource-intensive and constrains the number of descriptors and states that smaller projects can afford \cite{Prehofer2021}.\\

\noindent Current limitations in expert-elicitation approaches that inform CIB \cite{WEIMERJEHLE2006334}, MCDA, and the expert quantification of scenarios compound these constraints. These limitations fall into three areas: who participates, how decisions are made, and how storylines are translated into numbers. Scenario construction relies on small, purposively selected expert groups rather than representative samples, so consensus ratings can reflect majority views or group dynamics rather than the full range of expert or societal perspectives \cite{Weimer-Jehle2020,Guivarch2017,Sharmina2025}. The same applies to expert-led quantification and MCDA-based selection \cite{BAUER2017316,alcamo2008chapter,ishizaka2013multi}. The time and resources required can limit what information is gathered and how much it can be refined \cite{GrangerMorgan2001}. It is often difficult to trace how input decisions were reached \cite{burgman2015trusting}.\\ 

\noindent In addition, subjectivity in assessments and group dynamics can significantly influence outcomes, for example when consensus is determined by majority opinion or by the influence of dominant individuals, whilst dissenting views are rarely documented \cite{Granger2014,o2006uncertain,Trutnevyte2016}. The conversion from qualitative storylines to model parameters still relies heavily on modeller interpretation and can leave parameters only loosely tied to the narrative \cite{Guivarch2017,BRYANT201034}. Moreover, ensuring internal consistency across complex systems is cognitively demanding, whilst conventional panels often struggle to represent the full diversity of stakeholder views (e.g.\ consumer, industry, environmental) owing to logistical constraints, potentially leading to narrow or technocratic selection biases \cite{Weimer-Jehle2020}. Despite these limitations, expert elicitation remains essential for consistency, traceability, and making scenario logic explicit to decision-makers.\\

\noindent This paper presents a novel application of artificial intelligence (AI) to structured expert elicitation and decision-making to address and mitigate these challenges. In this proposed framework, a single large language model (LLM) simulates a multi-expert panel whenever expert consultation is required. It generates the required inputs for CIB (descriptors, states, and the cross-impact matrix), conducts a workshop that informs MCDA (e.g.\ criteria and weights for scenario selection), and quantifies the chosen scenarios (e.g.\ translating storylines into model parameters such as demand trajectories or carbon prices). Critically, the approach extends beyond scenario generation to decision support. The AI-simulated stakeholder panel performs MCDA, explicitly negotiating conflicting criteria (e.g.\ public acceptance versus infrastructure speed) to select pathways, thereby providing a transparent audit trail of the value judgements underpinning scenario selection. The CIB analysis implemented in this framework extends the prevailing workflow by incorporating confidence-coded judgement uncertainty and structural and dynamic shocks within a dynamic multi-period pathway framework, so that both robustness and shock-induced state switching are explored alongside the central workshop storyline (such aspects are often neglected in current scenario practice \cite{HANNA2021111984,McCollum2020}).\\ 

\noindent The approach delegates these key expert-elicitation and decision-making steps to AI-simulated panels, and offers documented reasoning at each step, a re-runnable protocol for iteration and sensitivity analysis, and lower time and resource intensity than convening conventional workshops. The design allows for facilitator or expert intervention at any stage, so that a hybrid workflow is possible. The pipeline can be run end-to-end without intervention or with selective oversight (e.g.\ reviewing or overriding individual descriptors or workshop phases). Parallel runs for ``scenario ensembles'' \cite{Guivarch2022} require minimal additional effort (but increase computational time). \\

\noindent The framework is demonstrated in a practical example that runs the scenario generation pipeline for Germany's energy transition to 2050, given European and global context. Germany aims for net-zero by 2045 and has set ambitious 2030 targets, but current progress is mixed \cite{UBA2025}. It is the focal country here, with power, buildings, transport, and industry in scope and European and global factors as drivers. Target region(s), time horizon, context, and panel expertise can be varied, making the approach suitable for a broader range of applications. The following section outlines the methodological pipeline in sequence.\\

\section{Methods}
\label{sec4}

\noindent The framework comprises four linked stages: an AI-led CIB workshop that elicits descriptors, states, and cross-impacts; a stochastic CIB analysis with dynamic pathways and shock mechanisms; an AI-led MCDA workshop for pathway selection; and a final quantification workshop that translates the selected pathway into model-ready inputs. Whilst the design is agnostic to the underlying LLM, the present implementation was run with Gemini 3 Flash Preview \cite{GeminiModels}.\\

\noindent The section is organised as follows. Section~\ref{1_workshop_CIB} describes the AI-led CIB workshop protocol and outputs. Section~\ref{2_cib_pycib} then specifies the CIB analysis, including uncertainty treatment, structural and dynamic shocks, and pathway simulation over the time grid. \ref{appA} details the AI-led MCDA workshop and pathway selection protocol, and \ref{appB} details the quantification workshop and translation of pathway states into numerical inputs.\\

\noindent It must be noted that CIB consistency in the impact-balance sense does not guarantee full narrative coherence. Given this, the quantified scenario remains conditional on the agreed workshop specification and on subsequent panel choices. In the present implementation, key stages, including quantification, are executed autonomously by AI panels; however, facilitator or expert intervention can be introduced at selected steps where closer alignment with particular assessment models or storyline is required. The framework nevertheless remains cost-effective and flexible, since panel composition, domain coverage, and study boundaries can be adjusted via prompts without additional data.\\

\noindent Beyond these implementation choices, the framework offers what may be termed a ``Glass Box'' validation standard. Full deliberative transcripts are retained (in the present application they exceed 250,000 words), thereby allowing a step-by-step audit of every decision, score, and quantification step and supporting intervention at each stage. Accompanying resources offer access to these transcripts. The following sections are organised to mirror the model workflow, and the transcripts are structured by workshop and phase (e.g.\ CIB phases 1-5, MCDA phases A0-E, quantification phases A-E), so that each subsection and phase can be traced in the accompanying materials. It must be stressed that this level of transparency is uncommon in scenario generation at present. The following subsection therefore begins with the elicitation design that provides the foundational inputs for all downstream stages.\\

\subsection{AI-led Cross-Impact Balance workshop}
\label{1_workshop_CIB}

\noindent The workshop follows a CIB design \cite{WEIMERJEHLE2006334} for eliciting plausible socio-technical futures for Germany's energy transition to 2050 within a European and global context. The central question is as follows: what consistent futures are plausible, given the interplay of policy, technology, infrastructure, society, and international drivers? The aim is to identify the factors that matter, define their possible states, and judge how they influence one another. The result of the workshop is a set of agreed outputs that support the construction of internally consistent scenarios and pathways for use in energy modelling and related analysis. The focus here is on the methodological design of the workshop and, in particular, on the representation of the panel of experts as a single AI system that simulates multiple domain experts\footnote{This concept is described, for example, in \cite{Thapen24}. Furthermore, a ``council-of-AIs'' configuration could also be used, in which different AI models act as individual experts rather than a single model simulating the whole panel.} and reaches consensus on each workshop item.\\

\noindent The workshop is organised in five phases, each of which builds on the agreed outcomes of the previous phase. The first phase establishes a common understanding of the workshop scope, horizon, and panel protocol. In the second phase, the panel selects which factors (descriptors) to include for the central question, up to a method-defined maximum of key factors. Descriptors may be reworded or clarified. This yields an agreed set of descriptors that drives all subsequent phases. In the third phase, for each agreed descriptor, the panel defines a small set of discrete, ordinal states (e.g.\ Low, Medium, High) with brief definitions.\\

\noindent This produces agreed state labels and definitions per descriptor. In the fourth phase, for each ordered pair of distinct descriptors (source-target), the panel agrees a cross-impact score and a confidence level for every combination of source state and target state (variant-level CIM); that is, for each cell in the source-state $\times$ target-state grid, the extent to which that source state supports or hinders that target state. The result is the full cross-impact matrix. In the fifth phase, the panel agrees the baseline scenario (initial state per descriptor), domain rules (e.g.\ forbidden state combinations), and dynamic elements (e.g.\ cyclic factors, threshold rules, time uncertainty). This delivers the full specification required for building CIB scenarios and pathways.\\

\noindent The AI panel comprises five experts with predefined domains of expertise. The first expert addresses policy and regulation: German and EU climate and energy policy, targets, carbon pricing, regulatory framework, and permitting. The second addresses technology and supply: renewables deployment, technology costs, storage, hydrogen, negative emission technologies, supply chains, and bioenergy. The third addresses infrastructure and markets: grid flexibility, European grid integration, electrification, industrial demand, fossil prices, the role of gas, and investment. The fourth addresses society and behaviour: public acceptance, consumer behaviour, land-use conflict, labour, and skills. The fifth addresses European and global context: EU policy alignment, international coordination, energy security pressure, and decarbonisation outcomes.\\ 

\noindent The number of experts on the AI panel, as well as their specific domains of expertise, can be adjusted as required to suit the needs of the application. This flexibility is illustrated in later sections of the modelled pipeline. The panel is given a common reference context, a limited shared baseline (scope, horizon, terminology). Experts are instructed to draw on their own domain knowledge and may extend or refine this baseline with evidence or framings from their respective fields.\\

\noindent A key feature of the approach is that the AI experts engage in elaborate discussions amongst themselves. They argue points, challenge each other's positions, and may disagree. The protocol explicitly instructs them to be critical and argumentative and not to converge too quickly. Dissent is expected and is noted before the group states the agreed outcome. For each workshop item, each expert must state their position and reasoning. The AI panel then deliberates, with experts drawing on their respective domains to question redundancy, scope, or alternative interpretations. Critically, this exchange of views and the possibility of disagreement are central to the method, since they surface different perspectives and evidence before a single conclusion is recorded. \\

\noindent The panel must nevertheless reach one consensus outcome per item and state why; that is, a short summary of the panel's reasoning is required. Consensus is based on this reasoned argumentation and collective deliberation. It draws on the common reference context and descriptor catalogue as a starting point, which experts may go beyond, and on the requirement to justify the consensus outcome. Where applicable, method constraints also apply (e.g.\ the maximum number of descriptors, or the score and confidence scales used in the cross-impact phase). The consensus result for each item is recorded as the workshop output for that item.\\

\noindent Procedurally, one workshop item is addressed at a time so that each decision receives full attention. Moreover, in later phases the items are processed in batches (e.g.\ groups of descriptors or source-target pairs) so that each step remains feasible in scope and time. This batching is a key methodological choice, since it ensures that the context window of the underlying AI is not exceeded. Between items or batches, context prompts are used to remind the panel of the current phase, the common reference context, and the task at hand, so that the deliberation remains coherent across the workshop. The workshop is implemented by a single AI system that simulates the five experts and their step-wise deliberation. The same protocol and common context are used throughout, whilst the specific task and context depend on the phase and the current item. A full workshop transcript is kept such that each decision, and its underlying reasoning, can be investigated manually if required.\\

\noindent In the present application, the AI experts select the main descriptors, pathways, and related variables autonomously on the basis of the common context and the central scenario question. The design can also be used in a more targeted way. For instance, the workshop prompt can be customised to guide the panel toward specific scenario families.\\

\subsection{Cross-Impact Balance analysis}
\label{2_cib_pycib}

\noindent At the core of the CIB method \cite{WEIMERJEHLE2006334} is a cross-impact matrix (CIM), $C$, the table of pairwise influence scores between descriptor states (recall: descriptors are factors, each with a set of discrete ordinal states) which is populated by workshop judgements. A scenario $z$ is one state chosen per descriptor (i.e.\ per factor), and $z_j$ denotes the state of descriptor $j$ in scenario $z$. For such a scenario, the impact score of target descriptor $j$ in state $l$ is

\begin{equation}
\label{eq:impact}
\theta_{j,l}(z) = \sum_{i \neq j} C_{i \to j}(z_i, l),
\end{equation}

where $i$ runs over descriptors, $z_i$ is the state of source descriptor $i$ in scenario $z$, $l$ denotes a possible state of descriptor $j$, and $C_{i \to j}(z_i, l)$ is the impact of source descriptor $i$ in its current state $z_i$ on target descriptor $j$ in state $l$. The impact balance for descriptor $j$ is the vector of these scores over all its states (one score per possible state of that descriptor). A scenario is consistent if, for every descriptor $j$, the chosen state $z_j$ attains the maximum (ties allowed) \cite{WEIMERJEHLE2006334}. That is, for each descriptor $j$ and for every state $l$ of that descriptor,

\begin{equation}
\theta_{j,z_j}(z) \ge \theta_{j,l}(z) \quad \text{for all } l.
\end{equation}

\noindent The analysis uses a fixed time horizon from 2025 to 2050. The time grid (or time periods) is the set of nodal years over this horizon; in the application reported here, $T = \{2025, 2030, ... , 2050\}$. An element $t \in T$ is referred to as a period; the index $t$ denotes a generic period throughout.\\

\noindent In standard (global) CIB succession, each iteration computes impact balances from the current scenario and updates descriptors simultaneously to their highest-scoring states (with deterministic tie handling), until a fixed-point or cyclic attractor is reached. In the extended analysis, this rule is augmented by uncertainty and shock terms and, where applicable, feasibility constraints on admissible successor states. Repeated updating in this manner eventually settles on a fixed scenario or a small cycle; these are the attractors implied by the cross-impact judgements. The foregoing defines standard CIB, which yields consistent scenarios and attractors from a fixed CIM. The analysis employed here extends this by using PyCIB \cite{PyCIB}, thereby adding confidence-coded and time-scaled judgement uncertainty together with structural and dynamic shocks for robustness.\\

\noindent Given the outputs from the workshop, a simulation-first probabilistic CIB approach is employed to generate dynamic multi-period pathways. Specifically, instead of relying on the enumeration of all CIB-consistent scenarios (which yields a static set and is feasible only for small spaces) or on the estimation of attractor weights alone (which fails to produce time-evolving pathways), a large number of independent Monte Carlo runs are performed over the time grid $T$. For each run, a CIM is sampled based on the workshop confidence levels. The system is subsequently evolved period by period via succession, cyclic transitions, and shocks. This process ensures that each run yields a single pathway (a linked sequence of scenarios over $T$).\\ 

\noindent This approach is selected because the scenario space is extensive. Moreover, the downstream MCDA workshop requires pathway-level outputs (linked timelines) to facilitate the selection of candidate futures. The system comprises the agreed 15 descriptors (three states each), and covers the time grid $T$. Judgement uncertainty is confidence-coded. That is, for each cell of the CIM the workshop agrees both a score $\mu$ (the point estimate for that cell) and a confidence code (1 to 5, where 5 denotes very high confidence). The confidence code determines a per-cell standard deviation $\sigma$, with higher confidence corresponding to smaller $\sigma$ (e.g.\ $\sigma = 0.2$ for confidence 5, $\sigma = 1.5$ for confidence 1). Each CIM cell is then sampled around its point estimate $\mu$ using the confidence-derived uncertainty scale $\sigma$, with support clipped to the CIB score range $[-3,+3]$; the implementation allows Gaussian or Student-$t$ draws, and in this application Student-$t$ (5 d.f.) is used for structural perturbations.\\

\noindent The workshop agreed to scale $\sigma$ by period using a multiplicative factor (e.g.\ 1.0 at 2025, increasing to 1.5 at 2050), so that uncertainty widens for more distant periods. A sampled matrix is drawn for each run (or per time step when $\sigma$ is scaled by period). Where dynamic shocks are present, the within-period succession produces a realised scenario under perturbed impact scores. The pathways stored for the subsequent MCDA workshop and for quantification are the realised scenarios at each period (the result of within-period succession under the sampled CIM, structural shocks where applied, and dynamic score perturbations). \\

\noindent Conventional expert panels can suffer from cognitive biases such as probability neglect regarding extreme events \cite{morgan1990uncertainty}.\footnote{Here, ‘extreme’ is used in a stress-testing sense, not as explicit modelling of disequilibrium regime shifts. In parts of the CIB literature, extremes refer either to rare high-impact events (e.g., extreme weather events) or to boundary scenarios spanning the solution space under a fixed matrix. In this application, structural and dynamic shocks are stochastic perturbations around elicited impacts; tail realisations may surface grey-swan-like pathway outcomes (plausible but often neglected) \cite{Taleb2007}, but the framework remains equilibrium/attractor-based and does not explicitly model non-equilibrium regime-break dynamics. This remains a limitation of current attractor-based CIB implementations and would require fundamental methodological extensions beyond standard CIB.} The shock mechanisms described below support exploration of robustness and pathway diversity that may receive less attention in conventional consensus processes (see also Results, Sect.~\ref{sec2}, and the footnote to Figure~\ref{fig:shocks}). As in the Results, two stochastic shock mechanisms are utilised. Structural shocks (random perturbations to the CIM entries) are re-sampled each period. This is done to stress-test the robustness of consistent scenarios when the world deviates from the elicited system. The workshop agreed to apply structural shocks in this application. The CIM is perturbed as

\begin{equation}
C \leftarrow C + \varepsilon, \qquad \varepsilon \sim \mathcal{N}(0,\Sigma),
\end{equation}

where $\varepsilon$ is a random perturbation matrix and $\Sigma$ is diagonal with diagonal entries $0.30^2$ (workshop-agreed scale for stress-testing) \cite{PyCIB}. The equation shows the Gaussian case; the implementation allows Gaussian or Student-$t$ perturbations, and in the application reported here the workshop chose Student-$t$ with 5 degrees of freedom. Robustness is reported as the fraction of runs in which a scenario remains consistent under the perturbed matrix.\\

\noindent Moreover, dynamic shocks (random additions to impact scores during succession) are applied at the impact-balance level during within-period succession (the iterative update within a single time step until the scenario stabilises) \cite{PyCIB}. They model time-varying disturbances that can nudge the balance between close competing states and occasionally induce switching. Persistence is introduced by an AR(1) (first-order autoregressive) structure \cite{PyCIB}. The perturbed score at period $t$ is

\begin{equation}
\theta'_{j,l}(t) = \theta_{j,l}(t) + \eta_{j,l}(t), \qquad \eta_{j,l}(t) = \rho \eta_{j,l}(t-1) + u_{j,l}(t),
\end{equation} 

\noindent Here $\theta'_{j,l}(t)$ is the perturbed impact score used during succession; $t \in T$ indexes the period; $\theta_{j,l}(t)$ is the (unperturbed) impact score~\eqref{eq:impact} evaluated at the scenario in period $t$ (i.e.\ $\theta_{j,l}(z(t))$ with $z(t)$ the scenario at period $t$). The term $\eta_{j,l}(t)$ is the dynamic shock (perturbation) added to the impact score for descriptor $j$ in state $l$ at period $t$. The coefficient $\rho$ is the AR(1) (first-order autoregressive) persistence coefficient and $u_{j,l}(t)$ are zero-mean innovations. In this application, Student-$t$ innovations with 5 degrees of freedom are used. The workshop agrees the long-run standard deviation $\tau$ of the AR(1) process. Assuming $|\rho|<1$ for stationarity, the innovation standard deviation is $\tau\sqrt{1-\rho^2}$.\footnote{For the AR(1) process $\eta_{j,l}(t) = \rho \eta_{j,l}(t-1) + u_{j,l}(t)$, the innovations $u_{j,l}(t)$ are zero-mean with standard deviation $\sigma_u$. The long-run variance of $\eta$ is $\sigma_u^2/(1-\rho^2)$ (standard for AR(1) processes). Setting the long-run standard deviation to the workshop-agreed $\tau$ gives $\tau^2 = \sigma_u^2/(1-\rho^2)$, so $\sigma_u = \tau\sqrt{1-\rho^2}$.} \\

\noindent Cyclic descriptors are factors whose state evolves exogenously between time steps. For each cyclic descriptor $j$, the panel agrees four parameters: the probability to stay in the current state (stay), to move one state along the ordinal scale (step), to move two states (step2), and an optional drift term (drift). Stay, step, and step2 are probabilities that sum to 1.0; drift is an optional parameter (typically 0). Between periods, the state of descriptor $j$ evolves according to the workshop-agreed transition probabilities. Specifically, $\mathbb{P}(z_j(t+1) = l' \mid z_j(t) = l)$, where $z_j(t)$ denotes the state of descriptor $j$ in period $t$, $l$ denotes the current state, and $l'$ denotes the successor state, is given by the agreed stay, step, step2, and drift values for descriptor $j$. Cyclic descriptors are locked during within-period succession so that $z_j(t+1)$ is determined by this transition law and not overwritten by the succession operator. Domain rules (forbidden pairs, implications) constrain feasibility. \\

\noindent The workshop specification also includes a threshold rule under which, when Public Acceptance is High and Permitting Pace is Fast, the impact of Permitting Pace on Grid Development is strengthened (add +1 to the Fast$\to$Strong cell). This reflects the judgement that favourable social acceptance and administrative speed together translate more effectively into grid build-out.\\

\noindent All inputs to the CIB analysis come from the workshop outlined in the previous subsection. They are the set of descriptors and their discrete states, the CIM (scores and confidence per source-target pair), the baseline scenario (initial state per descriptor in the first period, 2025), domain rules (forbidden pairs, implications), the time grid $T$, cyclic transition parameters (stay, step, step2, drift per cyclic descriptor, as defined above), and the final-phase options (structural and dynamic shock scales, cyclic parameters, and metadata such as endogenous/exogenous classification and mediated pathways for reporting). All of these are decided by the workshop participants. The PyCIB package \cite{PyCIB} can also compute diagnostics (e.g.\ near-miss rates, event-rate confidence intervals, and undersampling warnings) from the pathway and scenario outputs for use when low-probability or marginal-consistency behaviour is of interest; the present application does not report them in the main pipeline. The analysis thus produces a large number of dynamic pathway runs, from which candidate pathways are selected for the MCDA workshop (e.g.\ by frequency at 2050 and diversity of outcomes).\\

\noindent A general caveat applies here. CIB identifies scenarios that are internally consistent in the sense of impact balance (each descriptor is in the state that best fits the cross-impact judgements). That consistency does not, however, guarantee that the resulting storylines are free of narrative tensions \cite{Schweizer2020}. For example, the agreed descriptor set and cross-impacts may embed assumptions or framings that conflict with one another when read as a coherent narrative. The trajectory implied by a pathway may combine states that are individually consistent yet jointly difficult to justify in storyline form. Such tensions, where they arise, are carried forward in the pipeline. The candidate pathways passed to the MCDA workshop, and hence the pathway selected for quantification, inherit the narrative structure (and any latent tensions) of the CIB outputs.\\

\noindent The quantified scenario that emerges at the end of the pipeline is therefore conditional on both the CIB specification and the downstream choices made by the AI panels. In the current implementation the AI expert panels work through the pipeline autonomously, without facilitator or participant intervention. The results could therefore be improved upon at each step through targeted intervention by a facilitator or by domain experts (e.g.\ to validate or refine descriptors, cross-impacts, MCDA criteria and weights, or quantified assumptions). Users of the approach should remain aware that consistency in the CIB sense is a necessary but not sufficient condition for narrative coherence, and that checking or refining the storyline at selected stages (e.g.\ after CIB or after MCDA) may be advisable when the scenario is intended for communication or decision support.\\

\noindent Recall that \ref{appA} details the AI-led MCDA workshop and the pathway selection protocol, while \ref{appB} describes the quantification workshop and the process of translating pathway states into numerical inputs. The following section outlines selected results from each stage of the pipeline.

\section{Results}
\label{sec2}

\noindent The pipeline comprises four stages. Here, a pathway is a linked sequence of scenarios over the time horizon (one state per descriptor per period), and a descriptor is a factor with a set of discrete ordinal states (see Methods). The AI-panel CIB workshop establishes the scenario framework: factors (descriptors), their possible states, and the cross-impact matrix encoding how those states influence one another. The CIB analysis then produces an ensemble of internally consistent pathways over the time horizon. The MCDA workshop selects one pathway from that ensemble, with a documented rationale. The quantification workshop translates it into model-ready inputs: central trajectories and uncertainty ranges for key dimensions. Two figures summarise the results. Figure~\ref{fig:shocks} illustrates the shock mechanisms. Figure~\ref{fig:three_panel} presents results by stage: Panels (a) and (b) show CIB analysis; (c) and (d) show MCDA; and (e) to (g) show quantification. The subsections describe each stage in turn; the final subsection gives the pathway narrative, synthesis, and caveats.\\

\subsection{AI-panel CIB workshop and analysis}
\label{subsec1}

\noindent The AI-panel CIB workshop (Sect.~\ref{1_workshop_CIB}), run with five simulated experts with predefined domains of expertise (e.g. one for policy and regulation, one for technology and supply, and so on), produced the agreed descriptors, states, and cross-impact matrix. The workshop agreed on 15 descriptors, each with three ordinal states (e.g.\ Low, Medium, High, or method-specific labels such as Net-zero aligned, Medium emissions, High emissions) and brief definitions. The set includes, for example, Decarbonisation outcome (Germany's aggregate greenhouse gas emissions trajectory to 2050), Policy stringency (strength of climate and energy policy), Hydrogen role (extent of hydrogen's role in the energy system), and Renewables deployment (extent of renewables in the energy system). The full list and the pairwise cross-impact scores are given in the Methods and the accompanying resources. The CIB analysis (Sect.~\ref{2_cib_pycib}) takes those outputs and runs a Monte Carlo pathway simulation (many random runs over the time horizon), yielding a large ensemble of internally consistent pathways. \\    

\begin{figure}[htbp]
    \centerline{%
    \includegraphics[keepaspectratio]{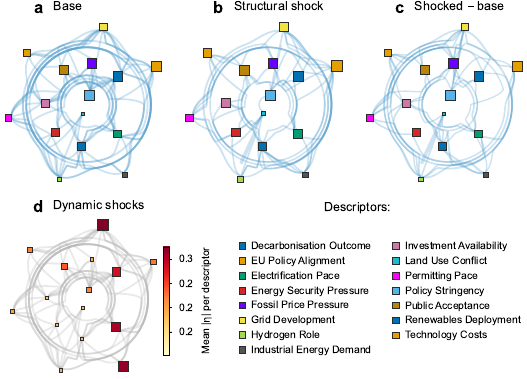}
    }
    \caption{Cross-impact matrix (CIM) as descriptor-level impact networks and the effect of the two stochastic shock mechanisms (structural and dynamic). Panels (a)-(c) share the same node-size scale (total outgoing impact strength). \textbf{a}, Base: workshop-agreed CIM; edge width is link strength (top 25\% by weight). \textbf{b}, Structural shock (standard deviation 0.30): same layout; edge width reflects shock magnitude, node size is total outgoing impact in the shocked CIM. \textbf{c}, Shocked - base: edge width reflects change in link weights (shocked CIM minus base); node size as in (a). \textbf{d}, Dynamic shocks (one period, 2040): node size and colour indicate mean absolute dynamic shock intensity per descriptor (scale not comparable to (a)-(c)). Data from synthetic AI workshops, CIM computed using PyCIB \cite{PyCIB}.}
    \label{fig:shocks}
    \end{figure}

\noindent The CIB analysis also uses two stochastic shock mechanisms to stress-test the robustness of consistent scenarios (formal definition in Methods Sect.~\ref{2_cib_pycib}). Figure~\ref{fig:shocks} illustrates them. Structural shocks perturb the workshop-agreed CIM and stress-test whether scenarios remain consistent when the influence table is perturbed away from the elicited central estimate. Panels (a)-(c) of the figure show the base CIM, the structural shock, and the difference (shocked minus base). Node size in (a)-(c) uses a common scale (total outgoing impact strength, i.e.\ how strongly each descriptor influences others), so the effect of the structural shock on descriptor strength is visible in (b), whilst (c) highlights which links changed most. \\

\noindent Dynamic shocks, as illustrated in Panel (d) of Figure~\ref{fig:shocks}, differ in that they do not alter the CIM. Instead, they introduce random perturbations to the impact scores when the succession operator updates each descriptor to its best-fitting state within each period. They essentially impose time-varying disturbances that can nudge the balance between closely competing states and occasionally induce state switching within a pathway, with persistence across periods given by an AR(1) (first-order autoregressive) structure (so that shocks in one period tend to carry over to the next). That is, structural shocks examine how robust scenarios are to errors in the elicited influence table, whilst dynamic shocks examine how small perturbations during succession can alter which scenario is realised, a stylised representation of the sensitivity of pathways to stochastic influences along the timeline. These mechanisms allow scenario robustness and pathway diversity to be explored.\\

\begin{figure}[htbp]
    \centerline{%
    \includegraphics[keepaspectratio]{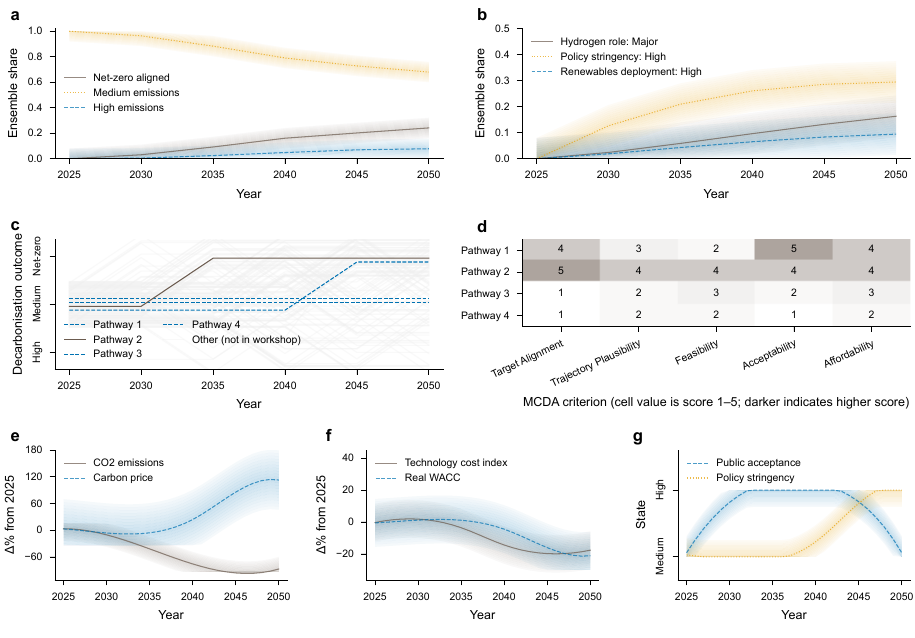}
    }
    \caption{Selected results of the AI-panel elicitation for socio-technical energy scenarios for Germany's energy transition. \textbf{a}, CIB: Decarbonisation outcome (all three states). \textbf{b}, CIB: one state each for Hydrogen role, Policy stringency, Renewables deployment. In (a)-(b), central lines are ensemble shares (the fraction of 10,000 Monte Carlo runs in which the descriptor is in that state at that time); bands are Wilson score confidence intervals (with a visual minimum-width fan). \textbf{c}, MCDA: Decarbonisation outcome for four candidate pathways; selected pathway highlighted, others and a CIB sample for context. \textbf{d}, MCDA: score matrix (pathways $\times$ criteria), scores 1-5. \textbf{e}, Quantification: CO$_2$ emissions and carbon price. \textbf{f}, Technology cost index and real WACC. In (e)-(f), central lines are trajectories as $\Delta$\% from 2025; bands are expert-elicited uncertainty ranges. \textbf{g}, Public acceptance and policy stringency (state over time); central trajectory smoothed (cubic spline), bands as in (e)-(f) mapped to state space. Underlying data are period-wise and step-wise; within-state variation not represented. Data from synthetic AI workshops.}
    \label{fig:three_panel}
    \end{figure}

\noindent The first two panels of Figure~\ref{fig:three_panel} summarise an excerpt of the scenario ensemble for a small set of descriptors chosen by AI experts. Panel (a) shows how the share of pathways in each of the three decarbonisation states changes over time for one descriptor, Decarbonisation outcome (Net-zero aligned, Medium emissions, High emissions). The shift towards net-zero-aligned outcomes by 2050 occurs because the experts' judgements imply that, as other factors (e.g.\ policy stringency, renewables deployment, grid development) evolve, the net-zero-aligned state receives stronger support, so more pathways end up there.\\ 

\noindent Panel (b) in Figure~\ref{fig:three_panel} shows how often selected states co-occur across the ensemble over time for three additional descriptors: Hydrogen role Major, Policy stringency High, and Renewables deployment High. The panels characterise the distribution of plausible futures and highlight the coherence between descriptors such as high policy stringency and high renewables deployment. For assessment modellers, this defines the range of decarbonisation and related assumptions that are internally consistent with the elicited system, so that scenario families can be delineated or assigned weights in sensitivity analyses.\\

\subsection{AI-panel MCDA workshop}
\label{subsec2}

\noindent From the scenario ensemble produced by the CIB analysis, the AI-panel MCDA workshop (\ref{appA}) first narrows to four candidate pathways for closer comparison, then, through deliberation, agrees on one. Each candidate is one possible realisation of the same rules and expert judgements. The selected pathway therefore has a specific sequence of internally consistent states over 2025-2050 that emerged in that realisation. In Figure~\ref{fig:three_panel}, panel (c) shows the four candidate pathways, with the selected pathway distinguished from the others. The trajectory in panel (c) moves from Medium emissions to Net-zero aligned (higher on the ordinal scale) because the panel chose this pathway as the preferred compromise between feasibility, ambition, and equity. For context, a sample of other CIB pathways not discussed in the workshop is also shown.\\

\noindent The AI-driven MCDA process captured distinct value conflicts during the selection of this trajectory. For instance, the simulated ``Industry'' persona (one of ten simulated stakeholder personas in this workshop) initially prioritised rapid grid expansion, whilst the ``Civil Society'' persona gave greater weight to public acceptance and land-use constraints. Through the deliberative protocol, the panel converged on a compromise pathway that balanced these competing objectives, making clear which trade-offs were accepted to reach a consensus. That choice is the bridge from many pathways to one, the single pathway passed on to the quantification workshop. In Figure~\ref{fig:three_panel}, panel (d) displays the score matrix (one row per pathway, one column per criterion) with the agreed scores from the same workshop. The per-persona weights and resulting ranking are recorded for audit. The matrix documents why that pathway was chosen and offers an audit trail, so that modellers can align sensitivity analyses and communication with the criteria that drove the choice (e.g.\ feasibility, ambition, equity) and the key personas behind it.\\

\subsection{AI-panel scenario quantification}
\label{subsec3}

\noindent The pathway selected in the MCDA step is then taken by the quantification stage (\ref{appB}), which, together with the associated narrative and descriptor definitions, translates it into forms needed by assessment models. The quantification turns the pathway's qualitative states (e.g.\ policy stringency Medium or High, public acceptance Low, Medium, or High) into numbers for use in models. Each state maps to a single numeric value per period. Because the pathway can stay in the same state for several periods, trajectories can exhibit flat segments and step changes rather than smooth curves, reflecting the discrete state-to-value mapping agreed in the workshop (i.e.\ the mapping from each descriptor state to a numerical value). Flat segments do not imply that real-world variables (e.g.\ carbon price) remain constant. They indicate that the pathway remained in the same qualitative state over those periods. In Figure~\ref{fig:three_panel} (panels e-g), the plotted trajectories are smoothed for display so that the curves appear as smooth transitions (e and f in $\Delta$\% from 2025, g in state space); the underlying quantified pathway remains period-wise and step-wise. Assessment modellers may interpolate between nodal years for their models if needed.\\

\noindent Panels (e)-(g) in Figure~\ref{fig:three_panel} show a small selection of the results. Panels (e) and (f) show central trajectories as $\Delta$\% from 2025 (emissions and carbon price; technology cost index and real weighted average cost of capital (WACC)). Panel (g) shows public acceptance and policy stringency as descriptor state over time (states present in the pathway), with the same smoothing and band convention. The bands in (e)-(g) are the uncertainty ranges (low and high) that the AI experts assigned to each variable (in (g), mapped to state space). They reflect deep uncertainty around the central trajectory and can be used directly as inputs or bounds in assessment models. The quantification step is run with three simulated expert personas covering integrated assessment models, energy system optimisation modelling, and data and input specialisation.\\ 

\noindent In this scenario pathway, as shown in panel (e), CO$_2$ emissions decline sharply as the system shifts from medium decarbonisation to a trajectory aligned with net-zero goals, following the policy stringency step-up from 2045. The baseline panel placed Decarbonisation outcome at Medium in 2025. Germany has cut emissions substantially and policy is oriented towards deep decarbonisation, so High emissions would overstate the baseline. Yet 2030 targets are at risk and net-zero by 2045 remains unsecured, so Net-zero aligned could not be justified. Policy stringency was likewise set at Medium, with ambition and instruments in place, but implementation gaps and political friction mean the full definition of High is not met for 2025. The CIM encodes an asymmetric feedback from decarbonisation outcome to policy stringency: the panel judged the evidence for policy tightening in response to poor outcomes to be stronger than the evidence for policy relaxation in response to good outcomes. Policy stringency stays at Medium through 2040 and, in this pathway realisation, steps to High in 2045 via the stochastic cyclic transition (as a cyclic descriptor; see panel (g)). The underlying emissions and carbon price mappings follow this path, flat through 2040 then stepping up in 2045 (smoothed in the figure for display).\\

\noindent Turning to panel (f), the technology cost index falls (in $\Delta$\% from 2025) as the pathway moves from Medium to Low technology costs, reflecting the cost reductions that accompany net-zero deployment. The baseline panel placed Technology Costs at Medium in 2025. Renewables are competitive in many segments, but hydrogen and some storage remain costly, so the aggregate points to Medium. Investment Availability was also set at Medium, with capital flowing into renewables but financing costs elevated and some segments constrained. The CIM encodes decarbonisation outcome as a proxy for deployment and learning. High emissions support high costs, net-zero aligned supports low costs. The pathway evolves to Low technology costs from 2040, consistent with the cross-impact support from the net-zero aligned outcome. The real WACC falls as investment availability moves from Medium to High, with the panel treating Investment Availability as the sole driver to avoid double-counting with other dimensions.\\

\noindent Panel (g) plots these two descriptors as state on the y-axis (the states present in the pathway; in this case Medium and High), with the same smoothing and uncertainty bands as (e) and (f). It traces public acceptance, which in this pathway realisation rises from 2025 to 2030, remains high through 2045, and steps down in the final period. The baseline panel placed it at Medium in 2025, weighing implementation and deployment friction over headline poll support. The drop in 2045-2050 is one possible outcome of the stochastic cyclic transition for public acceptance. Cyclic descriptors are factors whose state can change between periods according to agreed transition probabilities (rather than being fixed by the cross-impact logic alone), so that in any single pathway such a factor may or may not move from one period to the next. The panel set it as cyclic because it evolves between time steps due to discourse, events, local experience, and trust. In this pathway, acceptance steps down from High to Medium in the final period. This transition is independent of other descriptors; the concurrent policy stringency and carbon price step-up visible in panel (e) is a separate feature of this pathway realisation.\\

\subsection{Pathway narrative and synthesis}
\label{subsec4}

\noindent The narrative of the selected pathway is summarised as follows. It is a net-zero-aligned, hydrogen-limited pathway in which policy stringency and carbon price tighten late, coincident with the final push towards net-zero. The pathway traces Germany's energy transition from 2025 to 2050. The narrative and context are characterised by a decisive shift in 2035, consolidation through 2040-2045 (tighter policy and financing), and a deeply decarbonised end state with hydrogen limited throughout. ``Net-zero aligned'' is reflected in the agreed emissions level and in the combination of policy, technology cost, and price states. ``Medium emissions'' is reflected in the sustained level over 2025-2035. The experts' plausibility rules (e.g.\ coherence between policy stringency, renewables deployment, and decarbonisation outcome) are respected in the narrative.\\

\noindent Figure~\ref{fig:three_panel} traces selected descriptors and variables through the entire methodological pipeline, starting with the initial CIB workshop and analysis, continuing with the MCDA workshop where a single stakeholder-preferred pathway and its rationale are established, and concluding with the final quantification workshop that provides inputs for assessment models. This step-by-step visualisation underscores how broad ensembles of plausible futures are narrowed to a single, explicitly justified trajectory. A selection of end results is shown in panels (e)-(g), underpinned by the pathway narrative above and the selection rationale in panels (c) and (d). This supplies assessment modellers with the quantitative data and contextual information to run their models and interpret scenario assumptions in a consistent, traceable manner.\\

\noindent The following caveats close the results section. The trajectory is consistent with the experts' agreed qualitative states and the agreed mapping from those states to numbers. The discrete-state design can yield flat segments when the pathway stays in the same state across periods; this is a methodological artefact, not a claim about real-world constancy. In the figure (panels e-g), the step-wise trajectories are smoothed for readability (e and f as $\Delta$\% from 2025, g as state). Causal language in the storyline (e.g.\ that one factor drives or reinforces another) is interpretive and does not imply that the workshop structure establishes causation. The narrative is offered as a basis for quantification and for entry into assessment models (subject to the caveats on interpretation and causality). Moreover, the specific results are conditional on the present setup, namely an AI-expert panel given broad prompts and minimal steering towards particular scenario families. Closer guidance (e.g.\ model-specific input requirements or explicit storyline constraints), model-specific quantification, and a storyline aligned with current ambitions and policies would strengthen the pathway and its use in assessment models.\\

\section{Conclusions}
\label{sec3}

\noindent To the best of the authors’ knowledge, this analysis is the first of its kind to use an AI panel of experts for structured expert elicitation and decision-making within the context of socio-technical energy scenario building. Unlike generative AI applications that design or discover \cite[e.g.][]{Swanson2025}, this approach uses AI for elicitation, consensus-building, and decision-making within established methods. That is, no prior work is known to have applied such an AI panel within Cross-Impact Balance (CIB) or multi-criteria decision analysis (MCDA) settings to produce agreed multi-expert outputs for the energy transition or other domains. Moreover, the explicit consideration of stochastic structural and dynamic shocks that can stress-test scenario robustness and induce state switching, applied within dynamic multi-period pathway analysis in CIB, is novel to energy scenario generation and to CIB analysis itself. Thereby, it captures edge cases that might be smoothed over in a conventional consensus process.\\

\noindent Beyond the present application, the approach offers flexibility in scenario framing. In the implementation reported here, the AI experts select the main descriptors, pathways, and related variables autonomously, on the basis of the common context and a broadly defined main scenario question. The same design nevertheless admits a more directed use. The workshop prompts can be tailored so that the panel is steered towards a particular scenario family or storyline (e.g.\ an instance in which there is more focus on direct air capture and storage, or one in which demand reduction and sufficiency play a dominant role).\\

\noindent More generally, the approach makes high-quality socio-technical scenarios accessible to under-resourced regions (e.g.\ the Global South) by using synthetic expertise at near-zero marginal cost, reducing the dominance of well-funded institutions in scenario generation, and enabling analyses that previously lacked expert input. The framework can also produce a new, internally consistent scenario landscape in days instead of months, turning scenario planning from a slow, periodic exercise into a responsive tool that keeps assessment models updated despite rapid socio-economic and geopolitical change. Such a capability enables policymakers to explore how different panel compositions (e.g.\ NGOs versus industry) affect pathway construction and selection, acting as a ``wind tunnel'' for exploratory policy stress-testing that would be too costly and complex with conventional expert panels. That is, this capability effectively creates ``Virtual AI-Led Decision Laboratories'', as demonstrated by the MCDA workshop's configurable stakeholder personas. Assessment modellers also obtain trajectories and uncertainty ranges, together with the pathway narrative and MCDA rationale, for direct use in their models and for communicating scenario assumptions.\\

\noindent Given that MCDA and CIB are widely used and the constraints (cost, time, transparency) are common across expert-elicitation work, the AI-driven approach can be applied beyond energy to domains such as urban planning, land use, climate adaptation, healthcare, and environmental management. The approach is flexible. The number and domains of AI experts, research framework, and temporal or spatial boundaries can be changed via prompts without extra data, providing a cost-effective way to add AI expertise to expert elicitation and decision support. Parallel pipelines for scenario ensembles require minimal reconfiguration within the same framework, and CIB, MCDA, fuzzy cognitive maps, Delphi, Sustainable Development Goal assessment (e.g.\ using CIB to model goal interactions and their evolution over time), and similar methods can be attached at minimal additional cost to analyses that previously lacked such inputs.\\

\appendix
\section[Appendix A]{AI-led multi-criteria decision analysis workshop}
\label{appA}

\noindent This AI expert-led stakeholder workshop uses multi-criteria decision analysis (MCDA) to select one plausible and internally consistent pathway from a small set of candidate pathways produced by the CIB analysis outlined in the previous subsection. The workshop uses a simple additive weighting (SAW) approach: each pathway is scored on agreed criteria, persona-specific criterion weights are applied, and weighted totals are averaged across personas to rank pathways. The stakeholder elicitation protocol is method-agnostic. That is, elicited criteria, weights, and pathway scores can be post-processed with alternative MCDA aggregators (e.g., TOPSIS, VIKOR, PROMETHEE), subject to method-specific assumptions such as normalisation, criterion polarity, and preference-threshold settings.\\ 

\noindent One AI simulates multiple participants and reaches consensus on each item (Sect.~\ref{1_workshop_CIB}). Here the panel comprises ten stakeholder personas (policy and regulation, industry and utilities, civil society and NGOs, grid and infrastructure, research and advisory, consumer and households, finance and investors, labour and unions, regional and subnational, and environment and nature), distinct from the five CIB domain experts, and the task is pathway selection rather than CIB elicitation. \\

\noindent In the present application, candidate pathways from the CIB analysis are first narrowed using their frequency and diversity of outcomes across the ensemble; they are then screened for plausibility before stakeholder comparison. The screening excludes pathways with problematic temporal patterns, such as backsliding after improvement, implausible endpoint combinations, late-rush transitions, or large discontinuous shifts that lack a credible narrative. Given this, a sufficiently broad candidate pool is taken forward so that the panel can compare multiple plausible futures, including at least two pathways that reach the best outcome state at 2050. \\

\noindent The protocol follows that of the CIB workshop (Sect.~\ref{1_workshop_CIB}). The same structure applies, with ten stakeholder personas, and a full transcript is kept for audit. The workshop is organised in phases, each building on the previous. It draws on these candidate pathways. Each pathway is a timeline of internally consistent scenarios over $T$ (with the first period, 2025, not being in equilibrium). The application focuses on net-zero or close to net-zero pathways. The workshop is explicitly framed to prefer pathways that reach net-zero by 2050; status-quo or high-emissions pathways may be selected only if the panel justifies an exception.\\

\noindent In Phase A0 the panel reviews these candidates against the CIB descriptor and state definitions, flags any remaining narrative tensions (e.g.\ 2050 outcome-enabler inconsistency, large step-changes, or other storyline weaknesses), and selects four pathways for the full MCDA. Rejected pathways, together with the reasons for rejection, are documented. These four pathways form the candidate set for the remaining phases. \\

\noindent Phase A presents these four pathways to the panel for confirmation. In Phase B the panel agrees on a short list of evaluation criteria (e.g.\ feasibility, ambition, equity, cost, robustness, acceptability, with at least one criterion capturing target alignment such as net-zero alignment). In Phase C each persona proposes a weight vector over the criteria, with weights that are non-negative and sum to one; after discussion, one agreed vector per persona is recorded. In Phase D the panel agrees on one score per (pathway, criterion) on a common scale. In Phase E these persona-specific weights and agreed scores are combined to produce a pathway ranking, which then informs the final deliberative selection. \\

\noindent In Phase E, the panel determines an aggregate value $V_p$ for each pathway $p$ by combining the agreed score $s_{p,c}$ of pathway $p$ on criterion $c$ with the agreed weight $w_{r,c}$ assigned to that criterion by stakeholder persona $r$, and then aggregating across the $R$ stakeholder personas. That is, the final ranking reflects both the common score matrix and the plurality of stakeholder priorities. This is expressed as

\begin{equation}
\label{eq:mcda_value}
V_p = \frac{1}{R}\sum_{r=1}^{R}\sum_c w_{r,c} s_{p,c}, \qquad \sum_c w_{r,c} = 1 \text{ for each } r.
\end{equation}

\noindent The panel uses this ranking, together with deliberation, to agree on one selected pathway. That pathway is the full trajectory over $T$. The 2050 state is the selected scenario snapshot for single-period use if needed.\\

\noindent At the end of the workshop the primary output is one selected pathway (the full timeline over $T$), which is the result for later quantification and for use as the trajectory in assessment models. The 2050 scenario (the snapshot of the selected pathway at the end of the horizon) is also produced for backward compatibility or single-period use. The agreed criteria, per-persona weights, and pathway-criterion scores $s_{p,c}$ from Eq.~\ref{eq:mcda_value}, and hence the MCDA ranking, are recorded for audit and transparency. Because the panel composition (the ten stakeholder personas) is configurable, different compositions allow rapid exploration of how stakeholder priorities shape consensus. The method can thereby serve as a ``wind tunnel'' for exploratory policy stress-testing.\\

\noindent The MCDA workshop can be replaced by simpler, rule-based selection when a deliberative rationale is not required. A central pathway can be taken as the modal state per descriptor and period across the ensemble, or as the ensemble pathway closest to that modal trajectory. Outlier pathways can be added for stress-testing (e.g.\ by distance from the median or by extreme outcomes such as net-zero vs.\ high emissions). These approaches would avoid the MCDA workshop, reduce computational cost, and yield a compact statistical summary with explicit coverage of the tails, but they do not provide an audit trail of value trade-offs or support exploration of how stakeholder priorities affect the choice. The choice between MCDA and these alternatives therefore depends on whether the application needs a documented rationale for pathway selection or a statistical summary for robustness and sensitivity analysis.\\

\section[Appendix B]{AI-led scenario quantification}
\label{appB}

\noindent This AI expert-led workshop translates the selected pathway from the MCDA workshop (\ref{appA}) into quantitative inputs for assessment models. One AI simulates multiple participants and reaches consensus on each item (Sects.~\ref{1_workshop_CIB}, \ref{appA}). Here the panel comprises three experts and the task is quantification of the single selected pathway, rather than CIB elicitation or pathway selection. It must be noted that the automated quantification step does not merely generate text; rather, it produces input vectors for assessment models with explicit uncertainty ranges and thereby serves as a translator between storylines and the quantitative requirements of assessment models.\\

\noindent The quantification step is formulated in a general manner. The simulated AI experts first generate a list of variables that they identify as required for translating the selected pathway into model-ready form (e.g.\ demand trajectories, technology availability, or policy-related parameters). That list, and the subsequent assignment of numerical values or ranges to each variable, is produced autonomously by the panel on the basis of the pathway narrative and the common context. The same design can nevertheless be made more directed: a pre-specified list of inputs required by a particular energy or sectoral model can be supplied in the workshop context, so that the panel's task is to quantify those variables rather than to choose them. \\

\noindent Similarly, baseline values or ranges for key quantities can be supplied in advance, with the panel agreeing on deviations or refinements conditional on the selected pathway. The present application emphasises the autonomous case, in which the panel both identifies the relevant variables and agrees on their quantification. Applications that target a specific model would, however, benefit from a more model-specific setup and from clearer prior inputs on the key quantities or assumptions to be used as a reference.\\

\noindent The AI panel comprises three experts with predefined domains of expertise. The first addresses integrated assessment (long-term scenarios, emissions pathways, economy, energy, and land consistency, policy and macro drivers, and inputs such as carbon price, efficiency, and exogenous drivers). The second addresses energy system optimisation (capacity expansion, dispatch, renewables integration, grid, sector coupling, and model inputs such as capacity bounds, costs, and policy levers). The third addresses data and input specialisation (sectoral and technology data, uncertainty quantification and elicitation, economic input data, statistics, and ensuring defensible Low/Medium/High bands and plausible step changes). The panel is given the MCDA-selected pathway narrative and the CIB descriptor and state definitions (from Sect.~\ref{1_workshop_CIB}) as the authoritative reference for mapping narrative states (e.g.\ Low, Medium, High) to numbers.\\

\noindent As with the CIB and MCDA workshops, the quantification workshop is organised in phases, each building on the previous. Phase A presents the context and the selected pathway. The panel confirms scope and notes which model-input dimensions each expert considers most important (e.g.\ policy stringency, carbon price, capacity, costs). No agreed output block is required for this phase. In Phase B the panel agrees the list of model-input dimensions (variables and units, e.g.\ demand in petajoules per year (PJ/yr), carbon price in EUR/tCO$_2$, capacity in GW, technology costs in EUR/kW) and the translation matrix. For every dimension there is a numerical value for each descriptor state. Values may be time-independent or time-dependent per period. The result is an agreed set of dimensions and the full state-to-value matrix. \\

\noindent In Phase C the system applies the agreed matrix to the selected pathway to produce a baseline input table. The panel then reviews this baseline and proposes overrides only where needed (e.g.\ for plausibility or units). The result is the primary input table, i.e.\ the quantified pathway (Eq.~\ref{eq:quant_value}). Phase C may be run in batches over periods for context-window feasibility. The results are merged into one quantified pathway. Accounting identities (e.g.\ electrification share versus electricity demand, or sector demand versus total final energy) are not enforced during the quantification workshop; the panel agrees dimensions and overrides without an explicit identity step. An optional identity-enforcement step can be applied after the workshop to the merged quantified pathway, so that chosen identities hold. Enforcing such identities within the workshop would be a useful extension. After Phase C, a review phase allows the panel to confirm plausibility of the quantified pathway or to recommend revisions; if revisions are proposed, the system applies them and re-runs the review until the panel agrees. \\

\noindent Having confirmed plausibility in the review phase, in Phase D the panel first agrees uncertainty ranges per dimension and 2 to 4 extreme or bracket scenarios. In the present application these are single-period (2050) bounding cases constructed to span the envelope of plausible outcomes implied by the CIB analysis; this is a design choice, and the same protocol could be extended to produce multi-period extremes if required. They are selected along three axes: outcome-based (e.g.\ high emissions versus net-zero aligned), descriptor-based (stacking multiple adverse or favourable descriptor states), and frequency-based (low-frequency tail outcomes from the Monte Carlo ensemble). Each provides a complete set of quantitative values across all dimensions at 2050 and serves as a sensitivity bound for assessment modellers. The panel then agrees 2 to 3 alternative pathways as full timelines over the same dimensions and periods. In Phase E the panel consolidates the final package (quantified pathway, uncertainty ranges, extreme scenarios, alternative pathways) and agrees a model-ready checklist and schema summary. Phase E requires that plausibility has been confirmed in the review phase.\\

\noindent The quantification step can be stated formally. The selected pathway gives the state of each (relevant) descriptor in each period $t \in T$. For each dimension $d$, the panel agrees in Phase B which descriptor (or combination) drives it; denote by $s_d(t)$ the state (e.g.\ Low, Medium, High) of that descriptor in period $t$, and by $M_d(s)$ the agreed translation from state $s$ to a numerical value for dimension $d$ (the matrix may depend on period $t$ if time-dependent entries were agreed in Phase B). The quantified value for dimension $d$ in period $t$ is

\begin{equation}
\label{eq:quant_value}
x_d(t) = M_d(s_d(t)).
\end{equation}

\noindent When the matrix is time-dependent, $x_d(t) = M_d(s_d(t), t)$ is used instead. The set $\{x_d(t) : d \in D,\, t \in T\}$, where $D$ is the agreed set of dimensions and $T$ is the set of periods (the time grid defined in Sect.~\ref{2_cib_pycib}), is the primary input table produced in Phase C. The translation matrix and quantified pathway encode expert judgement on how to interpret narrative states and pathway trajectories.\\

\noindent At the end of the AI expert workshop the primary output is the quantified pathway (Eq.~\ref{eq:quant_value}), suitable for use as the trajectory in assessment models. The workshop also produces uncertainty ranges per dimension, 2 to 4 extreme scenarios (bounding-case snapshots at 2050 spanning the envelope of plausible CIB outcomes, for use in sensitivity analysis), and 2 to 3 alternative pathways as full timelines over all periods. The full transcript is kept for audit and transparency.\\

\noindent Quantification maps each period's descriptor state (e.g.\ Low, Medium, High) to a single value per dimension via an agreed translation matrix. When the pathway stays in the same state across periods the quantified series exhibits flat segments. When it changes state it exhibits step changes. In the present autonomous configuration the resulting trajectory can show sharp jumps (e.g.\ in carbon price or permitting lead time) that may be implausible for some modelling uses. Mitigation (tighter bands in the matrix, favouring smoother pathways, or post-processing the trajectory) is possible but each option introduces arbitrary choices. The present approach keeps the quantified pathway as the direct image of the agreed states and matrix, preserving traceability at the cost of accepting discrete, potentially volatile, period-to-period steps. Modellers may apply interpolation between nodal years for model-specific needs. \\


\section*{Data availability} 

\noindent Transcripts, data, and a full set of results are available at: \url{https://doi.org/10.5281/zenodo.19251041}. The code used to develop and/or analyse the models in this study is not publicly available, except for the components already provided within the PyCIB \cite{PyCIB} package. 

\section*{Acknowledgements} 

\noindent Special thanks to Fiona for her insightful discussions. AGR acknowledges funding by the Helmholtz Association under the programme ``Energy System Design''. Open Access is funded by the Deutsche Forschungsgemeinschaft (DFG, German Research Foundation) (491111487). The views and opinions expressed in this paper are those of the authors and do not necessarily reflect the official policy or position of any affiliated institutions, organisations, or funding bodies.

\renewcommand{\bibsection}{%
    \section*{References} 
    \addcontentsline{toc}{section}{References}
}
\bibliographystyle{elsarticle-num} 
\bibliography{references}

@article{Weimer-Jehle2020,
  author    = {Weimer-Jehle, Wolfgang and V{\"o}gele, Stefan and Hauser, Wolfgang and Kosow, Hannah and Poganietz, Witold-Roger and Prehofer, Sigrid},
  title     = {Socio-technical energy scenarios: state-of-the-art and CIB-based approaches},
  journal   = {Climatic Change},
  year      = {2020},
  volume    = {162},
  number    = {4},
  pages     = {1723--1741},
  doi       = {10.1007/s10584-020-02680-y},
}

@book{Taleb2007,
  author    = {Nassim Nicholas Taleb},
  title     = {The Black Swan: The Impact of the Highly Improbable},
  publisher = {Random House},
  year      = {2007},
  address   = {New York},
  isbn      = {978-1400063512}
}

@article{Swanson2025,
  author    = {Kyle Swanson and Wesley Wu and Nash L. Bulaong and John E. Pak and James Zou},
  title     = {The Virtual Lab of AI agents designs new SARS-CoV-2 nanobodies},
  journal   = {Nature},
  year      = {2025},
  month     = {October},
  volume    = {646},
  number    = {8085},
  pages     = {716--723},
  issn      = {1476-4687},
  doi       = {10.1038/s41586-025-09442-9}
}

@article{Sharmina2025,
  author    = {Maria Sharmina and Oliver Broad and John Barrett and Christian Brand and Alice Garvey and Harry Kennard and Jonathan Norman and James Price and Steve Pye and Jack Snape and Emily White},
  title     = {Policymaker-led scenarios and public dialogue facilitate energy demand analysis for net-zero futures},
  journal   = {Nature Energy},
  year      = {2025},
  volume    = {10},
  number    = {12},
  pages     = {1482--1492},
  doi       = {10.1038/s41560-025-01898-3},
  url       = {https://doi.org/10.1038/s41560-025-01898-3},
  issn      = {2058-7546}
}

@misc{UBA2025,
  author    = {{Umwelt Bundesamt}},
  year      = {2025},
  title     = {Treibhausgas-Projektionen 2025 -- Ergebnisse kompakt},
  url       = {https://www.uba.de/n115691en},
  note      = {Accessed: 2026-02-28}
}

@article{Schweizer2020,
  author    = {Vanesa J. Schweizer},
  title     = {Reflections on cross-impact balances, a systematic method constructing global socio-technical scenarios for climate change research},
  journal   = {Climatic Change},
  year      = {2020},
  volume    = {162},
  pages     = {1705--1722},
  doi       = {10.1007/s10584-019-02615-2},
}

@article{alcamo2008chapter,
  title={Chapter six the SAS approach: combining qualitative and quantitative knowledge in environmental scenarios},
  author={Alcamo, Joseph},
  journal={Developments in integrated environmental assessment},
  volume={2},
  pages={123--150},
  year={2008},
  publisher={Elsevier}
}

@book{morgan1990uncertainty,
  title={Uncertainty: a guide to dealing with uncertainty in quantitative risk and policy analysis},
  author={Morgan, Millett Granger and Henrion, Max and Small, Mitchell},
  year={1990},
  publisher={Cambridge university press},
  address={Cambridge, UK}
}

@article{BAUER2017316,
title = {Shared Socio-Economic Pathways of the Energy Sector -- Quantifying the Narratives},
journal = {Global Environmental Change},
volume = {42},
pages = {316-330},
year = {2017},
issn = {0959-3780},
doi = {10.1016/j.gloenvcha.2016.07.006},
author = {Nico Bauer and Katherine Calvin and Johannes Emmerling and Oliver Fricko and Shinichiro Fujimori and J{\'e}r{\^o}me Hilaire and Jiyong Eom and Volker Krey and Elmar Kriegler and Ioanna Mouratiadou and Harmen {Sytze de Boer} and Maarten {van den Berg} and Samuel Carrara and Vassilis Daioglou and Laurent Drouet and James E. Edmonds and David Gernaat and Petr Havlik and Nils Johnson and David Klein and Page Kyle and Giacomo Marangoni and Toshihiko Masui and Robert C. Pietzcker and Manfred Strubegger and Marshall Wise and Keywan Riahi and Detlef P. {van Vuuren}},
}

@article{Pregger2020,
  author    = {Thomas Pregger and Tobias Naegler and Wolfgang Weimer-Jehle and Sigrid Prehofer and Wolfgang Hauser},
  title     = {Moving towards socio-technical scenarios of the German energy transition---lessons learned from integrated energy scenario building},
  journal   = {Climatic Change},
  year      = {2020},
  volume    = {162},
  number    = {4},
  pages     = {1743--1762},
  doi       = {10.1007/s10584-019-02598-0},
}

@book{ishizaka2013multi,
  title={Multi-criteria decision analysis: methods and software},
  author={Ishizaka, Alessio and Nemery, Philippe},
  year={2013},
  publisher={John Wiley \& Sons},
  address={Chichester, UK}
}

@article{Trutnevyte2016,
  author    = {Evelina Trutnevyte and C{\'e}line Guivarch and Robert Lempert and Neil Strachan},
  title     = {Reinvigorating the scenario technique to expand uncertainty consideration},
  journal   = {Climatic Change},
  year      = {2016},
  volume    = {135},
  pages     = {373--379},
  doi       = {10.1007/s10584-015-1585-x},
}

@article{Guivarch2017,
  title     = {Scenario techniques for energy and environmental research: An overview of recent developments to broaden the capacity to deal with complexity and uncertainty},
  journal   = {Environmental Modelling \& Software},
  volume    = {97},
  pages     = {201--210},
  year      = {2017},
  issn      = {1364-8152},
  doi       = {10.1016/j.envsoft.2017.07.017},
  author    = {C{\'e}line Guivarch and Robert Lempert and Evelina Trutnevyte}
}

@article{WEIMERJEHLE2006334,
  title     = {Cross-impact balances: A system-theoretical approach to cross-impact analysis},
  journal   = {Technological Forecasting and Social Change},
  volume    = {73},
  number    = {4},
  pages     = {334-361},
  year      = {2006},
  issn      = {0040-1625},
  doi       = {10.1016/j.techfore.2005.06.005},
  author    = {Wolfgang Weimer-Jehle},
}

@article{KURNIAWAN2022102815,
title = {Towards participatory cross-impact balance analysis: Leveraging morphological analysis for data collection in energy transition scenario workshops},
journal = {Energy Research \& Social Science},
volume = {93},
pages = {102815},
year = {2022},
issn = {2214-6296},
doi = {10.1016/j.erss.2022.102815},
author = {J.H. Kurniawan and M. Apergi and L. Eicke and A. Goldthau and A. Lazurko and E. Nordemann and E. Schuch and A. Sharma and N. Siddhantakar and K. Veit and S. Weko}
}

@article{Prehofer2021,
  author    = {Preler, Tobias and Pregger, Thomas and V{\"o}gele, Stefan and Weimer-Jehle, Wolfgang},
  title     = {Linking qualitative scenarios with quantitative energy models: knowledge integration in different methodological designs},
  journal   = {Energy, Sustainability and Society},
  year      = {2021},
  volume    = {11},
  number    = {1},
  pages     = {25},
  doi       = {10.1186/s13705-021-00298-1},
}

@article{Guivarch2022,
  author    = {C{\'e}line Guivarch and Thomas Le Gallic and Nico Bauer and Panagiotis Fragkos and Daniel Huppmann and Marc Jaxa-Rozen and Ilkka Keppo and Elmar Kriegler and Tam{\'a}s Krisztin and Giacomo Marangoni and Steve Pye and Keywan Riahi and Roberto Schaeffer and Massimo Tavoni and Evelina Trutnevyte and Detlef van Vuuren and Fabian Wagner},
  title     = {Using large ensembles of climate change mitigation scenarios for robust insights},
  journal   = {Nature Climate Change},
  year      = {2022},
  volume    = {12},
  number    = {5},
  pages     = {428--435},
  doi       = {10.1038/s41558-022-01349-x}
}

@book{burgman2015trusting,
  title     = {Trusting judgements: How to get the best out of experts},
  author    = {Burgman, Mark A},
  year      = {2015},
  publisher = {Cambridge University Press},
  address   = {Cambridge, UK},
}

@article{GrangerMorgan2001,
  author    = {Granger Morgan, M. and Pitelka, Louis F. and Shevliakova, Elena},
  title     = {Elicitation of Expert Judgments of Climate Change Impacts on Forest Ecosystems},
  journal   = {Climatic Change},
  year      = {2001},
  volume    = {49},
  number    = {3},
  pages     = {279--307},
  doi       = {10.1023/A:1010651300697},
}

@book{o2006uncertain,
  title     = {Uncertain judgements: eliciting experts' probabilities},
  author    = {O'Hagan, Anthony and Buck, Caitlin E and Daneshkhah, Alireza and Eiser, J Richard and Garthwaite, Paul H and Jenkinson, David J and Oakley, Jeremy E and Rakow, Tim},
  year      = {2006},
  publisher = {John Wiley \& Sons},
  address   = {Chichester, UK},
}

@article{BRYANT201034,
  title     = {Thinking inside the box: A participatory, computer-assisted approach to scenario discovery},
  journal   = {Technological Forecasting and Social Change},
  volume    = {77},
  number    = {1},
  pages     = {34-49},
  year      = {2010},
  issn      = {0040-1625},
  doi       = {10.1016/j.techfore.2009.08.002},
  author    = {Benjamin P. Bryant and Robert J. Lempert},
}

@article{HANNA2021111984,
  title     = {How do energy systems model and scenario studies explicitly represent socio-economic, political and technological disruption and discontinuity? Implications for policy and practitioners},
  journal   = {Energy Policy},
  volume    = {149},
  pages     = {111984},
  year      = {2021},
  issn      = {0301-4215},
  doi       = {10.1016/j.enpol.2020.111984},
  author    = {Richard Hanna and Robert Gross},
}

@article{McCollum2020,
  author    = {McCollum, David L. and Gambhir, Ajay and Rogelj, Joeri and Wilson, Charlie},
  title     = {Energy modellers should explore extremes more systematically in scenarios},
  journal   = {Nature Energy},
  year      = {2020},
  volume    = {5},
  number    = {2},
  pages     = {104--107},
  doi       = {10.1038/s41560-020-0555-3},
}

@article{Granger2014,
  author    = {M. Granger Morgan},
  title     = {Use (and abuse) of expert elicitation in support of decision making for public policy},
  journal   = {Proceedings of the National Academy of Sciences},
  volume    = {111},
  number    = {20},
  pages     = {7176-7184},
  year      = {2014},
  doi       = {10.1073/pnas.1319946111},
}

@misc{PyCIB,
  author    = {Ross, Andrew G.},
  title     = {PyCIB Cross-Impact Balance (CIB) Analysis Package},
  year      = {2026},
  doi       = {10.5281/zenodo.18367511},
  url       = {https://github.com/ag-ross/PyCIB},
  note      = {Software},
}

@misc{Thapen24,
  author    = {Thapen, Nick},
  title     = {Better LLM Prompting using the Panel-of-Experts},
  year      = {2024},
  url       = {https://www.sourcery.ai/blog/panel-of-experts/},
}

@misc{GeminiModels,
  author    = {{Google}},
  title     = {Models: {Gemini} {API}},
  year      = {2025},
  url       = {https://ai.google.dev/gemini-api/docs/models},
  note      = {Google AI for Developers},
}

\end{document}